\documentclass[10pt, a4paper]{article} 
\usepackage{amsmath,mathtools}
\usepackage{amssymb}
\usepackage{dsfont}
\usepackage{graphicx,psfrag,epsf, subfigure}
\usepackage{enumerate}
\usepackage[backref,breaklinks,colorlinks,citecolor=blue]{hyperref}
\usepackage{natbib}
\usepackage{caption}
\usepackage{tabls}
\usepackage{pdflscape}
\usepackage{algorithm}
\usepackage{algorithmic}
\usepackage{afterpage}
\usepackage{listings}
\usepackage{authblk}
\newtheorem{theorem}{Theorem}[section]

\newenvironment{remark}[1][Remark]{\begin{trivlist}
        \item[\hskip \labelsep {\bfseries #1}]}{\end{trivlist}}

\newcommand{\T}{\intercal}
\title{A fast algorithm for computing distance correlation}
\begin{document}
\date{}
\author[1]{Arin Chaudhuri \thanks{arin.chaudhuri@sas.com}}
\author[1]{Wenhao Hu \thanks{wenhao.hu@sas.com}}
\affil[1]{Internet of Things, SAS Institute Inc.}
\maketitle

\begin{abstract}
 Classical dependence measures such as Pearson correlation, Spearman's $\rho$,
 and Kendall's $\tau$ can detect only monotonic or linear dependence. To overcome
 these limitations, \cite{szekely2007measuring} proposed distance covariance
 as a weighted $L_2$ distance between the joint characteristic function and
 the product of marginal distributions. The distance covariance is $0$ if and
 only if two random vectors ${X}$ and ${Y}$ are independent. This measure has
 the power to detect the presence of a dependence structure when the sample
 size is large enough. They further showed that the sample distance covariance
 can be calculated simply from modified Euclidean distances, which typically
 requires $\mathcal{O}(n^2)$ cost. The quadratic computing time greatly limits
 the application of distance covariance to large data. In this paper, we present
 a simple exact $\mathcal{O}(n\log(n))$ algorithm to calculate the sample
 distance covariance between two univariate random variables. The proposed
 method essentially consists of two sorting steps, so it is easy to implement.
 Empirical results show that the proposed algorithm is significantly faster than
 state-of-the-art methods. The algorithm's speed will enable researchers to explore
 complicated dependence structures in large datasets.
\end{abstract}

\noindent%
\vfill

\section{Introduction}
Detecting dependencies between two random vectors $X$ and $Y$ is a fundamental
problem in statistics and machine learning. Dependence measures such
as Pearson's correlation, Spearman's $\rho$, and Kendall's $\tau$ are
used in almost all quantitative areas; example areas are bioinformatics
\citep{guo2014inferring, sferra2017phylo_dcor} and time-series \citep{zhou2012measuring}.
However, those classical dependence measures are
usually designed to detect one specific dependence structure such as a monotonic
or linear structure. It is easy to construct highly dependent $X$ and $Y$
whose dependence cannot be detected by classical dependence measures. To overcome
these limitations, \cite{szekely2007measuring, szekely2009brownian} proposed distance covariance
as a weighted $L_2$ distance between the joint characteristic function and the
product of marginal characteristic distributions. The distance covariance is
$0$ if and only if two random vectors ${X}$ and ${Y}$ are independent. A
closely related measure is the Hilbert-Schmidt independence
measure (HSIC). HSIC has been extensively studied in machine learning
literature \citep{gretton2005measuring, gretton2008kernel, pfister2018kernel}.
\cite{sejdinovic2013equivalence} established equivalence of distance covariance
with HSIC.

Despite the power of sample distance covariance to detect a dependence structure, its use 
for large sample sizes, is inhibited by the high
computational cost required. The sample distance covariance and HSIC computation
typically requires $\mathcal{O}(n^2)$ pairwise distance (kernel) calculations
and $\mathcal{O}(n^2)$ memory for storing them. This is undesirable and greatly
limits the application of distance correlation to large datasets. In the era of big
data, it is not rare to see data that consists of millions of observations.
For such data, an $\mathcal{O}(n^2)$ algorithm is almost impossible to run on a
personal computer. To approximate the distance covariance or HSIC for large
data, Nystr\"om approach or the random Fourier feature method is often
adopted. However, the use of these approximations leads to a reduction in power
\citep{zhang2018large}. In this article, we describe an exact method to compute
the sample distance covariance between two univariate random variables with
computational cost $\mathcal{O}(n\log(n))$ and memory cost $\mathcal{O}(n)$. Our
proposed method essentially consists of just two sorting steps, which makes it
easy to implement.

A closely related $\mathcal{O}(n\log(n))$ algorithm for sample distance
covariance was proposed by \cite{huo2016fast}. Our algorithm differs from
\cite{huo2016fast} in the following ways: First, 
they implicitly assume that there are no ties in the data (see Algorithm
1 and proof in \cite{huo2016fast}), whereas our proposed method is valid for any
pair of real-valued univariate variables. In practice, it is common to see
datasets with ties, especially for discrete variables or bootstrap sample. 
Second, we use a merge sort instead
of an AVL tree-type implementation to compute the Frobenius inner product of
the distance matrices of $\mathbf{x}$ and $\mathbf{y}$. Empirical results show
that our proposed method is significantly faster; for example, for one million
observations our MATLAB implementation runs 10 times faster (finishing in 4
seconds) on our desktop, whereas the implementation in \cite{huo2016fast}
requires 40 seconds. Because our implementation consists only of MATLAB code
while the key step in the \cite{huo2016fast} routine is implemented in C, even
greater speed increases are possible by rewriting the critical parts of our
implementation in C.
 
The rest of paper is organized as follows. In Section \ref{sec:pre},
we briefly introduce the definition of distance covariance and
its sample estimate. In Section \ref{sec:method}, we describe 
the proposed $\mathcal{O}(n\log(n))$ algorithm for sample distance covariance.
In Section \ref{sec:sim}, experiment results are presented. 
Finally, conclusions and remarks are made in Section \ref{sec:conc}.  

\section{Some Preliminaries}
\label{sec:pre}
Denote the joint characteristic function of $X \in \mathbb{R}^p$ and $Y \in \mathbb{R}^q$ 
as $f_{X, Y}(t, s)$, and denote the
marginal characteristic functions of $X$ and $Y$ as  $f_{X}(t)$ and  $f_{Y}(s)$,
respectively. Denote $|\cdot|_k$ as the Euclidean norm in $\mathbb{R}^k$. 
The squared distance covariance is defined as the weighted $L_2$ distance between
 $f_{X, Y}(t, s)$ and $f_{X}(t)\cdot f_{Y}(s)$,
\begin{align*}
\mathcal{V}^2(X,Y) = \int_{\mathbb{R}^{p+q}}|f_{X, Y}(t, s) - f_{X}(t)f_{Y}(s)|^2w(t, s)d t d s,
\end{align*} 
where $w(t,s) = (c_pc_q|t|^{1+p}_p |s|^{1+q}_q)^{-1}$, $c_p$ and $c_q$ are constants.
It is obvious that $\mathcal{V}^2(X,Y) = 0$ if and only if $X$ and $Y$
are independent. Under some mild conditions, the squared distance covariance 
can be defined equivalently as the expectation of Euclidean distances
\begin{align}
\label{eq:distCov}
\mathcal{V}^2(X,Y)  = & E(|X - X'|_p|Y - Y'|_q) - 2 E(|X - X'|_p|Y - Y''|_q) \nonumber \\
   & + E(|X - X'|_p)E(|Y - Y'|_q), 
\end{align}
where $(X, Y), (X', Y'),$ and $(X'', Y'')$ are identical independent copies from the joint
distribution of $(X,Y)$.

The squared distance correlation is defined by
\begin{align}
\label{eq:distCor}
\mathcal{R}^2(X, Y) = 
\begin{cases*}
\frac{\mathcal{V}^2(X, Y)}{\sqrt{\mathcal{V}^2(X, X) \mathcal{V}^2(Y, Y)}}, & \text{if  $\mathcal{V}^2(X, X) \mathcal{V}^2(Y, Y) > 0$ } \\ 
0 & \text{otherwise}.
\end{cases*}
\end{align}

Let $\mathbf{X} = (x_1, \dots, x_n)^\T$ and $\mathbf{Y}=(y_1,\dots,y_n)^\T$
be the sample collected. Define
\begin{align*}
    a_{ij} = |x_i - x_j|_p, & \ & b_{ij} = |y_i - y_j|_q\\
    a_{i.} = \sum_{j=1}^{n} a_{ij}, & \  & b_{i.} = \sum_{j=1}^{n} b_{ij}\\
    a_{..} = \sum_{i=1}^{n} a_{i.}, & \  & b_{..} = \sum_{i=1}^{n} b_{i.}\\
    D = \sum_{1 \leq  i,j \leq n} a_{ij} b_{ij}
\end{align*}
The squared sample distance covariance between $\mathbf{X}$ and $\mathbf{Y}$ is 

\begin{equation}
    \label{eq:1}
    \begin{split}
    \MoveEqLeft
        \mathcal{V}^2_n(\mathbf{X}, \mathbf{Y}) = \frac{D}{n^2} - \frac{2}{n^3}\sum_{i=1}^{n} a_{i.} b_{i.} + \frac{a_{..}b_{..}}{n^4},
    \end{split}
\end{equation}
which is similar in form to \eqref{eq:distCov}. 
The squared sample distance correlation is given by
\begin{align}
\label{eq:distCor2}
\mathcal{R}_n^2(X, Y) = 
\begin{cases*}
\frac{\mathcal{V}_n^2(\mathbf{X}, \mathbf{Y})}{\sqrt{\mathcal{V}_n^2(\mathbf{X}, \mathbf{X}) \mathcal{V}_n^2(\mathbf{Y}, \mathbf{Y})}}, & \text{if  $\mathcal{V}_n^2(\mathbf{X}, \mathbf{X})
    \mathcal{V}_n^2(\mathbf{Y}, \mathbf{Y}) > 0$ } \\ 
0 & \text{otherwise}.
\end{cases*}
\end{align}
From \eqref{eq:1}, it is easy to see a $\mathcal{O}(n^2)$ brute force algorithm
exists for distance covariance. However, the brute force implementation is difficult to handle large datasets. 
Moreover, the p-value of
distance covariance or correlation is typically calculated by using permutation test,
which makes it more computationally intensive.

If we can compute $D$ and \emph{all} $a_{i.},b_{i.}$ for $1 \leq i \leq
n$ and $D$ in $\mathcal{O}(n \log n)$ steps, then we can also compute
$\mathcal{V}^2_n(\mathbf{X}, \mathbf{Y})$ in $\mathcal{O}(n \log n)$ steps.

In this paper, we consider the case where $X$ and $Y$ are univariate random
variables; that is, $p = q = 1$. 
For the rest of this document we assume that
$x_1 \leq x_2 \leq \dots \leq x_n$ (because after an $\mathcal{O}(n \log n)$ sort
step, we can ensure that $x_1 \leq x_2 \leq \dots \leq x_n$).

\section{Fast Algorithm for Distance Covariance}
\label{sec:method}
Define the function $I(x)$ as
\begin{equation}
    I(x) =
    \begin{cases*}
        1 & if $x > 0$ \\
        0 & otherwise.
    \end{cases*}
\end{equation}
For any two real $x$ and $y$ we have
\begin{equation}
    \label{eq:2}
    |x - y| = (x-y)(2I(x-y) - 1).
\end{equation}
We use \eqref{eq:2} extensively in the rest of paper.

\subsection{Fast computation of the $a_{i.}$ and $b_{i.}$}
\label{subsec:aibi} 

Define $s_i = \sum_{j=1}^{i} x_i$ for $1 \leq i \leq n$ and note that $s_1,
\dots, s_n$ can be computed in $\mathcal{O}(n)$ time.

Since $x_1 \leq x_2 \leq \dots \leq x_n$ we have
\begin{align}
    a_{i.} &= \sum_{j < i} (x_i - x_j) + \sum_{j > i}(x_j - x_i) \nonumber \\
    &= (2i-n)x_i + (s_n - 2s_i).
    \label{eq:aiCal} 
\end{align}
So $a_{1.},\dots, a_{n.}$ can be computed in $\mathcal{O}(n)$ time.

We can use an $\mathcal{O}(n\log n)$ sorting algorithm to determine a permutation $\pi(1),$ $\pi(2),$ $\dots,$ $\pi(n)$ of
$1,2,\dots,n$ such that $y_{\pi(1)} \leq y_{\pi(2)} \leq \dots \leq y_{\pi(n)}$.
Therefore as in \eqref{eq:aiCal},  $b_{\pi(1)}, \ldots, b_{\pi(n)}$ can be computed in $\mathcal{O}(n)$ time after $y_1, \ldots, y_n$ is sorted. 

\subsection{Fast computation of D}
In this subsection, we describe an $\mathcal{O}(n\log(n))$
algorithm for computing $D$.
First, we have 
\begin{align}
    \label{eq:D}
    D &= \sum_{i=1}^{n}\sum_{j=1}^{n} |x_i - x_j| | y_i - y_j| \nonumber \nonumber \\
    &= 2 \sum_{i=1}^{n}\sum_{1 \leq j < i} |x_i - x_j| |y_i - y_j|. 
\end{align}
In \eqref{eq:D} note that $|x_i - x_j| = x_i - x_j$ if $1 \leq j \leq i$,
thus showing that
\begin{align}
    \label{eq:D1}
    \frac{D}{2} =& \sum_{i=1}^n\sum_{1 \leq j < i} (x_i - x_j) (y_i - y_j)(2I(y_i - y_j) - 1) \nonumber \\
      =&  2\sum_{i=1}^n\sum_{1 \leq j < i} (x_i - x_j)(y_i - y_j) I(y_i - y_j) \nonumber\\ 
       & \quad -\sum_{i=1}^{n}\sum_{1 \leq j < i}
    (x_i - x_j)(y_i - y_j).
\end{align}

Let $m_x = \sum_{i=1}^n x_i / n$ and $m_y = \sum_{i=1}^n y_i / n$.  
Now the second term in \eqref{eq:D1} is 
\begin{align} 
    \sum_{i=1}^{n}\sum_{1 \leq j < i} (x_i - x_j)(y_i - y_j) &= \frac{1}{2} \sum_{ 1 \leq i,j \leq n} (x_i - x_j)
(y_i - y_j) \nonumber \\
    &= n \sum_{i=1}^n (x_i - m_x) (y_i - m_y) \nonumber .
\end{align}
Therefore it can be computed in $\mathcal{O}(n)$ steps.

Define $U_i = \{j : 1 \leq j < i, \  y_j < y_i\}$ for $ 1 \leq i \leq n.$
The first term in \eqref{eq:D1} can be easily expressed in terms of $U_i$.
Note that
\begin{align}
& \sum_{i=1}^n\sum_{1 \leq j < i} (x_i - x_j)(y_i - y_j) I(y_i - y_j) \nonumber \\
& = \sum_{i=1}^n x_i y_i\sum_{j \in U_i} 1 - \sum_{i=1}^n x_i \sum_{j \in U_i} y_j
- \sum_{i=1}^n y_i \sum_{j \in U_i} x_j + \sum_{i=1}^n  \sum_{j \in U_i} x_j y_j.
\end{align}
Thus the first term in \eqref{eq:D1} is expanded into a sum of four terms,
each of which is of the form
\begin{equation}
    \begin{split}
    \sum_{i=1}^n s_i \sum_{j \in U_i} t_j  .
    \end{split}
\end{equation}

For any $t_1, \dots, t_n$, define $d_i = \sum_{j \in U_i} t_j .$ If it can be
shown that \emph{all} of $d_1,\dots,d_n$ can be computed in $\mathcal{O}(n \log
n)$ time, then the sample distance covariance can be computed in $\mathcal{O}(n \log
n)$ time. We will show this in the next subsection. The preceding arguments lead to the
following theorem.

\begin{theorem}
	For any real-valued univariate variables with sample \linebreak $\mathbf{x}= (x_1, \dots, x_n)^\T$ and 
       $\mathbf{y}=(y_1,\dots,y_n)^\T$, the sample distance covariance
       $\mathcal{V}_n(\mathbf{x}, \mathbf{y})$
	can be computed in $\mathcal{O}(n \log n)$ time. 
	\label{thm:time}
\end{theorem}

\subsection{Fast computation of $d_i$ where $d_i = \sum_{j \in U_i}t_j$}

We are given a series $y_1,\dots,y_n$ along with weights $t_1,\dots,t_n$.
Define $U_i = \{ j: 1 \leq j < i, \ y_j < y_i \}$.
Our objective is to compute $d_1,d_2,\dots,d_n$ in $\mathcal{O}(n \log n)$ steps where 
$$
d_i = \sum_{ \substack{ j < i \\ y_j < y_i} } t_j = \sum_{j \in U_i} t_j
$$
for $1 \leq i \leq n.$

It is well known that the number of inversions in a permutation can be obtained
by a merge sort \citep{Ginat:2004:SCS:1026487.1008020}. 
We use a similar strategy
to compute the $d_i$ while performing a merge sort on $y_1,\dots,y_n$ to sort the $y_i$ in
decreasing order.

Merge sort works by successively merging sorted subarrays until the final array
is sorted. Assume that we also keep an auxiliary array for storing the original
indices of each element of the array, another auxiliary array for storing
partially computed $d_i$ (say $d[1],\dots,d[n]$), and third one for storing the
partial sums of the intermediate results.

A merge sort makes $\lceil \log_2 n \rceil$ passes over the data. At
the beginning of the $k^{\text{th}}$ pass, the array consists of $\lceil
\frac{n}{2^{k-1}} \rceil$ contiguous subarrays, each of which is sorted in
decreasing order. We merge two consecutive subarrays $\lfloor \frac{n}{2^k}
\rfloor$ times.

Let $A= [y_{n_1},y_{n_2},\dots,y_{n_r}]$ and $B=[y_{m_1},y_{m_2},\dots,y_{m_s}]$
be two such consecutive subarrays with $y_{n_1} \geq y_{n_2} \geq \dots \geq
y_{n_r}$ and $y_{m_1} \geq \dots \geq y_{m_s}$. We can also assume $n_i
< m_j$ for all meaningful $i,j$ (here $m_i$ and $n_j$ refer to the original
indices of these elements).
During the merge step, if we notice that $y_{m_{\alpha}} > y_{n_\beta}$ for some
$\alpha,\beta$, then $y_{n_{\beta}},y_{n_{\beta+1}},\dots,y_{n_r}$ are all the
terms in $A$ that are less than $y_{m_\alpha}$ and we increment $d[m_\alpha]$ by
$t_{n_{\beta}} +\dots+t_{n_r}$. Note that if we also store running sums as we
output the results, $d[m_{\alpha}]$ can be computed using just one difference.

The extra computation of maintaining the additional auxiliary arrays does
not increase the order of computation. The detailed algorithm is presented in
Algorithm \ref{alg:fastD}. For a better understanding of the proposed algorithm,
we also include MATLAB code in the Appendix.

\begin{algorithm}
	\caption{Fast computation of $d_1, \ldots, d_n$}
	\label{alg:fastD}
	\hspace*{\algorithmicindent} \textbf{Input:} $Y = (y_1, y_2, \ldots, y_n)$
	                                         and $T= (t_1, t_2, \ldots, t_n)$\\
	\hspace*{\algorithmicindent} \textbf{Output:} $d = (d_1, \ldots, d_n)$
	
	\begin{algorithmic}[1]
		\STATE Initialize $idx$ to be a $2 \times n$ matrix whose first row is $(1, 2, \ldots, n)$ and whose second row
        is $(0, 0, \ldots, 0)$
		\STATE $i = 1;\ r = 1;\ s = 2;$
		\WHILE{$i <n$}
		
		\STATE $gap = 2*i;\ k = 0;\ idx\_r = idx[r, ];$
		\STATE $csumT = cusum(T[idx\_r]);\ csumT = (0,\ csumT); /*\text{cum sum}*/$  		
		\STATE $j = 1;$
		\WHILE{$j < n$}
		\STATE $st1 = j;\ e1 = \min(st1 + i - 1,n);$
		\STATE $st2 = j + i; e2 = min(st2 + i - 1,n);$
		
		\WHILE{$st1 <= e1$ and $st2 <= e2$}
		\STATE $k = k + 1$
		\STATE $idx1 = idx\_r[st1];\ idx2 = idx\_r[st2];$
		\IF{$Y[idx1] >= Y[idx2]$}
		\STATE $idx[s,k] = idx1;$ $st1 = st1 + 1;$
		\ELSE
		\STATE $idx[s,k] = idx2;\ st2 = st2 + 1;$
        \STATE  $d[idx2] = d[idx2] + (csumT[e1+1] - csumT[st1]);$
		\ENDIF
		\ENDWHILE
		
		\IF{$st1 <= e1$}
		\STATE $kf = k + e1 - st1 + 1;$
        \STATE $idx[s,(k+1):kf] = idx\_r[,st1:e1];$             
        \STATE $k = kf;$
        
        \ELSIF{$st2 <= e2$}
        \STATE $kf = k + e2 - st2 + 1;$
        \STATE $idx[s,(k+1):kf] = idx\_r[,st2:e2];$
        \STATE $k = kf;$
		\ENDIF
		
		\STATE $j = j + gap;$
		\ENDWHILE
		
		\STATE $i = gap;$
		\STATE $r = 3 - r;\ s = 3 - s;$
		\ENDWHILE
		
	  \RETURN d
	\end{algorithmic}

\end{algorithm}

	\begin{table}
		\centering
		\caption{Illustration of fast algorithm for series 
			$Y = (3, 5, 7, 3, 8, 4, 6, 7)^\T$ and 
			weight $T = (1, 5, 3, 2, 4, 6, 7, 5)^\T$. The index row denotes
			the original order of series, and the cusumT row denotes the cumulative sum
			of $t_i$. The proposed 
			algorithm finishes in three iterations. 
			Final results : $(d_1, \ldots, d_8) = (0, 1, 6, 0, 11, 3, 14, 21)$.
		}
		\label{tab:ex1}
		\begin{tabular}{cccccccccc}
			\hline
			& index  & 1 & 2 & 3 & 4 & 5 & 6 & 7 & 8 \\
			& y      & 3 & 5 & 7 & 3 & 8 & 4 & 6 & 7 \\
			Iter 0	& t      & 1 & 5 & 3 & 2 & 4 & 6 & 7 & 5 \\
			& cusumT & 1 & 6 & 9 & 11 & 15 & 21 & 28 & 33 \\
			&  d     & 0 & 0 & 0 & 0 & 0 & 0 & 0 & 0 \\
			\hline
			& index & 2 & 1 & 3 & 4 & 5 & 6 & 8 & 7  \\
			& y     & 5 & 3 & 7 & 3 & 8 & 4 & 7 & 6  \\
			Iter 1  & t     & 5 & 1 & 3 & 2 & 4 & 6 & 5 & 7  \\
			& cusumT & 5 & 6 & 9 & 11 & 15 & 21 & 26 & 33 \\
			& d     & 1 & 0 & 0 & 0 & 0 & 0 & 7 & 0  \\
			\hline
			& index & 3 & 2 & 1 & 4 & 5 & 8 & 7 & 6  \\
			& y     & 7 & 5 & 3 & 3 & 8 & 7 & 6 & 4  \\
			Iter 2 & t     & 3 & 5 & 1 & 2 & 4 & 5 & 7 & 6  \\
			& cusumT & 3 & 8 & 9 & 11 & 15 & 20 & 27 & 33 \\
			& d     & 6 & 1 & 0 & 0 & 0 & 13 & 6 & 0  \\
			\hline
			& index & 5 & 3 & 8 & 7 & 2 & 6 & 1 & 4  \\
			Iter 3	& y     & 8 & 7 & 7 & 6 & 5 & 4 & 3 & 3  \\
			& d     & 11 & 6 & 21 & 14 & 1 & 3& 0 & 0  \\
			\hline
		\end{tabular}
	\end{table}

We illustrate the proposed algorithm by using a simple series, \newline $Y = (3, 5, 7,
3, 8, 4, 6, 7)^\T$ and weight $T = (1, 5, 3, 2, 4, 6, 7, 5)^\T$. The iteration
history is presented in \autoref{tab:ex1}. The index row in \autoref{tab:ex1}
denotes the original order of the $Y$ series, and the cusumT row denotes the
cumulative sum of $t_i$. The algorithm finishes in $\log_2(8) = 3$ iterations
and outputs the results $(d_1, \ldots, d_8) = (0, 1, 6, 0, 11, 3, 14, 21)$.
Consider the computation of $d_7$ and $d_8$ for example. At iteration 1, we merge
$y_7,y_8$, because $y_7 < y_8$ we exchange $y_7$ and $y_8$ and set $d_7 = 0$ and
$d_8 = 7$. At iteration 2, we merge $y_5, y_6$ and $y_8, y_7$; and because $y_8 > y_6$ and
$y_8 < y_5$ we increment $d_8 = 7 + t_6 = 13$ and because $y_7 > y_6$ we increment $d_7
= 0 + t_6 = 6$. At the final merge, we note $y_8 > y_2$ so we increment $d_8 = 13 + t_2 + t_1 +
t_4 = 13 + (11 - 3) = 21$, where $t_2 + t_1 + t_4$ is calculated from the difference
of cumulative sums of $t_i$. Similarly for $d_7$ we have $d_7 = 6 + t_2 + t_1 + t_4 = 6 +
(11 - 3)= 14.$

The proposed algorithm has the same order of computational cost as a merge sort.
Denote the computational cost as $\mathcal{T}(n)$. We know
$\mathcal{T}(n) = 2 \mathcal{T}(n/2) + \mathcal{O}(n)$.
It is trivial to show that the complexity is
$\mathcal{O}(n\log(n))$ by using the master theorem \citep{cormen2001introduction}.
As a byproduct of calculating all $d_i$, $y_1, y_2, \ldots, y_n$ are sorted.
Therefore, calculating all $b_{i.}$ as in Section \ref{subsec:aibi} takes only
extra $\mathcal{O}(n)$ time.

\begin{remark}
As discussed previously, the proposed algorithm essentially consists of two
sorting steps. First we sort X and calculate $a_{i\cdot}$ for $i = 1, \ldots, n$.
Then we sort Y and calculate $D$ and all $b_{i\cdot}$ for $i = 1, \ldots, n$.
The correctness of the described algorithm can be easily concluded from the previous
discussion. To verify the correctness of our implementation we matched our
numbers with a simple brute force $\mathcal{O}(n^2)$ implementation and we
confirmed that the numbers match exactly.

We also note another factor that impacts performance. A non recursive merge
sort algorithm makes $\lceil \log_2(n) \rceil$ passes over the data, and in the
$k^{\text{th}}$ pass makes $\approx \frac{n}{2^k}$ merges of two subarrays of
size $2^{k-1}$. If $n$ is large then for small $k$ we end up merging a large
number of small subarrays, which has a large overhead. We have observed that 
the speed increases by a factor of $\approx 1.3$ if we replace these
initial merges by a single insert sort step --- that is, we divide the input array
into $\approx \frac{n}{16}$ groups of length $16$ and sort each one of them
using insert sort, store the intermediate $d$ values, and then continue
with the merge steps with $k \geq 5$. A MATLAB implementation is provided
as supplemental material to this paper.
\end{remark}

\section{Experiments}
\label{sec:sim}
In this section, we compare the speed of the proposed fast algorithm
with the dyadic updating method \citep{huo2016fast} and also with the
brute force implementation. We implemented the
proposed algorithm and the brute force method in MATLAB. 
For the dyadic updating method, we use the
authors' MATLAB implementation, in which the key step is implemented
in C. Therefore, this comparison strongly favors the 
dyadic updating method because more than $90\%$ of 
its calculations are done in C instead of MATLAB according to the MATLAB code profiler.
All simulations 
are run on a PC with Intel\textsuperscript{\textregistered} 
Xeon\textsuperscript{\textregistered} Gold CPU @ 2.40GHZ processor
and 16GB memory running MATLAB version 9.0.0.341360 (R2016a).

The data are generated from a simple quadratic model, 
$y_i = x_i^2 + \epsilon_i$ for $i = 1, \ldots, n$, where $x_i$ and $\epsilon_i$
are i.i.d. standard normal. For each sample size $n = 2^6,\ 2^8,\ 2^{10},\ 2^{12},\ \ldots,
2^{22}$, average running time is calculated based on 10 generated samples. 
The speed comparison results are presented in \autoref{tab:time}. The column
Merge Sort 2 in the table corresponds to proposed algorithm with 
initial merge steps replaced by insertion sort. 
For a moderately sized data (for example $n = 2^{14}$),
the dyadic update method is about 25 times faster
than the brute force implementation, whereas
our proposed algorithm is about 200 times faster.
Since brute force implementation requires 
$\mathcal{O}(n^2)$ space, no results are available 
when $n \geq 2^{16}$. 
For large data ($n \geq 2^{16}$), proposed method is 7 to 10 times
faster than dyadic updating. Greater speed increases
are expected if our proposed algorithm is implemented in C. 
Note that the \emph{p}-value
of the distance covariance is typically obtained by a permutation
test, which is computationally intensive. For $1,000$ permutations and
$n = 2^{14} = 16,384$,
the brute force implementation takes around $9.5 \times 1000$ seconds 
($158$ minutes) versus $0.048 \times 1000$ seconds (0.8 minutes) or 
 $0.024 \times 1000$ seconds (0.4 minutes) for proposed $\mathcal{O}(n\log(n))$ implementation.

{\tablinesep=2.5ex\tabcolsep=10pt
\begin{table}[!h]
	\caption{Average time in seconds for 10 replications. The numbers
	in parentheses are the estimated standard deviation of running time.
    No results are available for brute force implementation when 
    $n \geq 2^{16} = 65536$
    because the required memory exceeds the maximum memory size. Column Merge Sort 2 corresponds to proposed algorithm with 
    initial merge steps replaced by insertion sort. } 
   \label{tab:time}
	\centering
	\begin{tabular}{lcccc}
		\hline
	n	& Brute Force & Dyadic Update & Merge Sort & Merge Sort 2 \\
	  \hline
	 $2^6$ & 0.0011(0.0052) &	0.0006(0.0001)	& 0.0004(0.0002)  &0.0001(0.0001)\\
	$2^{8}$  & 0.0011(0.0003)	& 0.0027(0.0013) & 0.0011(0.0004) & 0.0004(0.0001)\\
    $2^{10}$  & 0.0416(0.0034)  & 0.0107(0.0007) & 0.0031(0.0009) & 0.0013(0.0003)\\
    $2^{12}$  & 0.6014(0.0160)	& 0.0678(0.0115) & 0.0126(0.0026) & 0.0059(0.0022)\\
    $2^{14}$ & 9.5096(0.1134)	& 0.3585(0.0247) &	0.0479(0.0054) & 0.0241(0.0036)\\    
   	
    $2^{16}$   &   -   & 1.7816(0.0757)	& 0.2511(0.0190) & 0.1624(0.0261)\\
    $2^{18}$  &   -   & 9.2188(0.2154)	& 1.1844(0.0552) & 0.8258(0.0625)\\
    $2^{20}$ &   -   & 45.302(0.4065)	& 5.4248(0.1077) & 4.0098(0.1737)\\
    $2^{22}$ &   -   & 219.04(2.7388)	& 25.345(0.4245) & 19.8976(0.4665)\\    
\hline	
	\end{tabular}
\end{table}
}

\section{Conclusions}
\label{sec:conc}
In this paper, we presented an $\mathcal{O}(n\log(n))$ algorithm
for calculating the sample distance covariance for univariate variables. 
For multivariate random variables, random projection approach can be
adopted \citep{huang2017statistically}, which depends on calculation
of distance covariance for univariate variables.
The proposed algorithm is intuitive and 
simple to implement. Empirical results show that it 
outperforms existing methods in the literature. Our algorithm will
speed up any computational technique that depends on the calculation of univariate distance covariance for example,
feature screening \citep{li2012feature}.

The proposed faster algorithm provides a tool for scientists to explore complicated dependence structures using
distance covariance in larger data than what was previously possible.

\newpage
\nocite{*}
\bibliography{dcov}
\bibliographystyle{apa}

\newpage
\bigskip
\begin{center}
	{\large\bf Appendix}
\end{center}
\section*{MATLAB Code for Fast Distance Covariance}
\definecolor{mygreen}{rgb}{0,0.6,0}
\lstset{
    basicstyle=\footnotesize,
    commentstyle=\color{mygreen},  
    escapeinside={\%*}{*)},
    keywordstyle=\color{blue}
}

\begin{lstlisting}[language=Octave]
function covsq = fastDcov(x,y)
%fastDcov computes distance correlation between column vectors x and y

%Sort the data by x. This does not change the answer.
n = length(x);
[x, Index] = sort(x);
y = y(Index);

%a_x is the vector of row sums of distance matrix of x
si = cumsum(x);
s = si(n);
a_x = (-(n-2):2:n).' .*x + (s - 2*si);

%Compute Frobenius inner product between the
%distance matrices while finding indexes corresponding
%to sorted y using merge-sort.

%Weight vectors
v = [x y x.*y];
nw = size(v, 2);

% The columns of idx are buffers to store sort indices and output buffer
% of merge-sort
% we alternate between them and avoid uneccessary copying
idx = zeros(n, 2);
idx(:,1) = 1:n;

% iv1, iv2, iv3, iv4 are used to store sum of weights.
% On output, for %*$\color{mygreen} 1 \leq j \leq n $ *) 
% %*$\color{mygreen} \text{iv1}(j)=\displaystyle\sum_{\substack{i<j,\\y_i<y_j}}1$*)
% %*$\color{mygreen} \text{iv2}(j)=\displaystyle\sum_{\substack{i<j,\\y_i<y_j}}x_i$*)
% %*$\color{mygreen} \text{iv3}(j)=\displaystyle\sum_{\substack{i<j,\\y_i<y_j}}y_i$*)
% %*$\color{mygreen} \text{iv4}(j)=\displaystyle\sum_{\substack{i<j,\\y_i<y_j}}x_i y_i$*) 
iv1 = zeros(n,1);iv2 = zeros(n,1);iv3 = zeros(n,1);iv4 = zeros(n,1);
% The merge sort loop.
i = 1; r = 1; s = 2;
while i < n;
    gap = 2*i;
    k = 0;
    idx_r = idx(:,r);
    csumv = [zeros(1, nw); cumsum(v(idx_r,:))];
    for j = 1:gap:n;
        st1 = j; e1 = min(st1 + i - 1,n);
        st2 = j + i; e2 = min(st2 + i - 1,n);
        
        while (st1 <= e1) && (st2 <= e2);
            k = k +1;
            idx1 = idx_r(st1);
            idx2 = idx_r(st2);
            
            if y(idx1) >= y(idx2);
                idx(k,s) = idx1;
                st1 = st1 + 1;
            else
                idx(k,s) = idx2;
                st2 = st2 + 1;
                
                iv1(idx2, 1) = iv1(idx2) + e1 -st1 +1;
                iv2(idx2) = iv2(idx2) + (csumv(e1+1, 1) - csumv(st1, 1));
                iv3(idx2) = iv3(idx2) + (csumv(e1+1, 2) - csumv(st1, 2));
                iv4(idx2) = iv4(idx2) + (csumv(e1+1, 3) - csumv(st1, 3));
            end;
          
        end;        
        if st1 <= e1;
            kf = k + e1 - st1 + 1;
            idx((k+1):kf, s) = idx_r(st1:e1,:);
            k = kf;
        elseif st2 <=e2;
            kf = k + e2 - st2 + 1;
            idx((k+1):kf, s) = idx_r(st2:e2,:);
            k = kf;
        end;      
    end;
    
    i = gap;
    r = 3-r; s = 3-s;
end;
% d is the Frobenius inner product of the distance matrices
covterm =  n*(x - mean(x)).' * (y - mean(y));

c1 = iv1.' * v(:, 3);
c2 = sum(iv4);
c3 = iv2.' * y;
c4 = iv3.' * x;

d = 4*((c1 + c2) - (c3 + c4)) - 2*covterm;

% b_y is the vector of row sums of distance matrix of y
ySorted = y(idx(n:-1:1, r));
si = cumsum(ySorted);
s = si(n);
b_y = zeros(n, 1);
b_y(idx(n:-1:1, r)) = (-(n-2):2:n).' .*ySorted + (s - 2*si);

%covsq equals %*$\color{mygreen}\mathcal{V}_n^2(x,y)$*) the square of the distance covariance
%between x and y.
nsq = n*n;
ncb = nsq*n;
nq = ncb*n;
term1 = d / nsq;
term2 = 2* (a_x.' * b_y) / ncb;
term3 = sum(a_x) * sum(b_y) / nq;

covsq = (term1 + term3) - term2;

\end{lstlisting}
\end{document}